\documentclass[preprint,showpacs,aps,10pt,twocolumn,tightenlines]{revtex4}%
\usepackage{amsfonts}
\usepackage{amsmath}
\usepackage{amssymb}
\usepackage[tcidvi]{graphicx}%
\begin{document}
\preprint{ }
\title[ ]{Temperature and field dependence of Dynamics in the Electron-Glass}
\author{Z. Ovadyahu}
\affiliation{The Hebrew University, Jerusalem 91904, Israel }

\begin{abstract}
We describe several experimental methods to quantify dynamics in electron
glasses and illustrate their use in the glassy phase of crystalline
indium-oxide films. These methods are applied to study the dependence of
dynamics on temperature and on non-ohmic electric fields at liquid helium
temperatures. It is shown that over a certain range of temperature the
dynamics becomes slower with temperature or upon increasing an applied
non-ohmic field, a behavior suggestive of a quantum-glass. It is demonstarted
that non-ohmic fields produce qualitatively similar results as raising the
system temperature. Quantitatively however, their effect may differ marekdly.
The experimental advantages of using fields to mimic higher temperature are
pointed out and illustrated.

\end{abstract}
\maketitle

\section{ Introduction}

Glasses are characterized by a slow approach to equilibrium, a process often
described as a journey through a phase space composed of many local
energy-minima that are separated by barriers \cite{1}. The motion through this
phase space is hindered by these barriers; the system is trapped for
relatively long time in one of the local minima before it manages to cross a
barrier and move to another metastable state. At finite temperatures, crossing
a barrier is usually assumed to be controlled by thermal activation.
Alternatively, the barriers may be crossed by quantum mechanical tunneling,
and this will become the dominant mechanism at sufficiently low temperatures.
It should be possible to tell the quantum-mechanical regime from the thermally
activated regime by the dependence of the glassy dynamics on temperature $T$.
Dynamics in the classical glass speeds up as the temperature increases, and
this feature has been studied extensively in many systems. The case in the
quantum scenario is less clear. It is commonly believed that, as long as the
barrier does not change, tunneling is temperature independent. However, in a
dissipative environment, which seems a relevant consideration for the
situation at hand \cite{2}, tunneling rate $\gamma$ may depend on temperature;
$\gamma$ may increase or decrease with $T$ depending on the interplay between
the barrier parameters, the coupling of the tunneling object to the
environment, and the temperature range \cite{3}.

In this paper we describe several methods to probe the dynamics in electron
glasses. These are used to study the dynamics in the glassy phase of
indium-oxide films as function of temperature, non-ohmic electric field $F$,
and disorder. A previous work, based on the the aging behavior, showed that
the dynamics in this system does not slow down as $T$ is lowered \cite{4}. The
more sensitive methods used here demonstrate that over some range of
temperatures and sample parameters, the dynamics may actually \textit{slow
down} with temperature. Similar result is observed when instead of increasing
the sample temperature the excess energy of the electrons is increased by
applying non-ohmic electric fields. It is demonstrated that increasing the
field affects various quantities in the same direction as raising the
temperature although quantitatively their effect may be quite different. The
experimental advantages of using non-ohmic fields to mimic higher temperatures
conditions are utilized to illustrate some unique features of the electron glass.

\section{Experimental}

\subsection{Sample preparation and measurement techniques}

Several batches of $In_{2}O_{3-x}$ samples were used in this study. All were
thin films of crystalline indium-oxide (30-50\AA \ thick, with lateral
dimensions of $\simeq$1x1 mm). These were e-gun evaporated on 110$\mu$m cover
glass. The sheet resistance R$_{\square}$ of the samples was limited to the
range 1M$\Omega$-2G$\Omega$ at $T=4.11K$. This range includes samples with
large enough disorder to exhibit significant glassy effects, and small enough
sheet resistances to allow reasonably high frequencies for the ac conductance
measurements (to facilitate a reasonable temporal resolution). Gold film was
evaporated on the backside of the glass substrate and served as a gate
electrode (configuring the sample as a field-effect device). Conductivity of
the samples was measured using a two terminal ac technique employing a
1211-ITHACO current preamplifier and a PAR-124A lock-in amplifier. The
measurements as function of non-ohmic fields were performed with the samples
immersed in liquid helium at $T=4.11K$ held by a 100 liters storage-dewar.
Measurements as function of temperatures were made in a $^{3}$He refrigerator
with the samples in vacuum attached to a copper stage along with a
Ge-thermometer. For these measurements, the ac voltage bias was small enough
to ensure ohmic conditions. Temperature was controlled with Oxford Instruments
ITC503 employing a RuO thermometer. Fuller details of sample preparation,
characterization, and measurements techniques are given elsewhere \cite{5}.

\subsection{Procedures for measuring dynamics}

The main goal of this work was to characterize the dynamics in the electron
glass and its dependence on external parameters in particular the effect of
temperature, and electric fields. The natural (history free) relaxation law of
the electron glass is logarithmic, a behavior that may extend over more than 5
decades in time \cite{6}. Following e.g., a quench-cooling from high
temperature to $\approx4K,$ the sample conductance relaxes slowly, and depends
on time as $G(0)-a\log(t)~$(where $G(0)$ is the conductance at $T=4K$
immediately after the cool-down). Having no characteristic time scale, such a
relaxation law does not allow for a unique definition of a relaxation time
$\tau$. It is possible however to give an empirical scheme that yields a
measure of dynamics. In the following we describe three such methods.

\subsubsection{The two-dip-experiment}

The first example is the two-dip-experiment \cite{5}, which involves the
following procedure. First, the sample is allowed to equilibrate at the
measuring temperature with a voltage $V_{g}^{o}$ held at the gate. Then, a
$G(V_{g})$ trace is taken by sweeping $V_{g}$ across a voltage range
straddling $V_{g}^{o}$. The resulting $G(V_{g})$ exhibits a minimum centered
at $V_{g}^{o}$ which reflects an inherent feature of a hopping system - its
equilibrium conductance is at a local minimum \cite{7}. At the end of this
sweep, a new gate voltage, $V_{g}^{n}$, is applied and maintained at the gate
between subsequent $V_{g}$ sweeps that are taken consecutively at latter times
(measured from the moment $V_{g}^{n}$ was first applied). Each such sweep
reveals two minima; One at $V_{g}^{o}$ which fades away with time, and the
other at $V_{g}^{n}$ whose magnitude increases with time. The
two-dip-experiment then amounts to studying the dynamics of the "forming"\ of
a cusp at a newly imposed $V_{g}^{n}$ and the "healing"\ of an `old' cusp at
$V_{g}^{o}$. A characteristic-time $\tau$ is defined as the time at which the
amplitude of the cusp at $V_{g}^{n}$ just equals the amplitude of the cusp at
$V_{g}^{o}$. This procedure is experimentally well-defined, and it is fairly
independent of the particular relaxation law. The two-dip-experiment has been
used before to study the dependence of the dynamics on carrier concentration,
and on magnetic field \cite{7}.

\subsubsection{The double-conductance-excitation method}

The main method we use in this work to quantify dynamics is a variation on the
two-dip-experiment. The procedure is illustrated in figure 1.%
\begin{figure}
[ptb]
\begin{center}
\includegraphics[
trim=0.000000in 4.673947in 0.000000in 1.871037in,
natheight=14.583300in,
natwidth=9.791400in,
height=2.6161in,
width=3.1808in
]%
{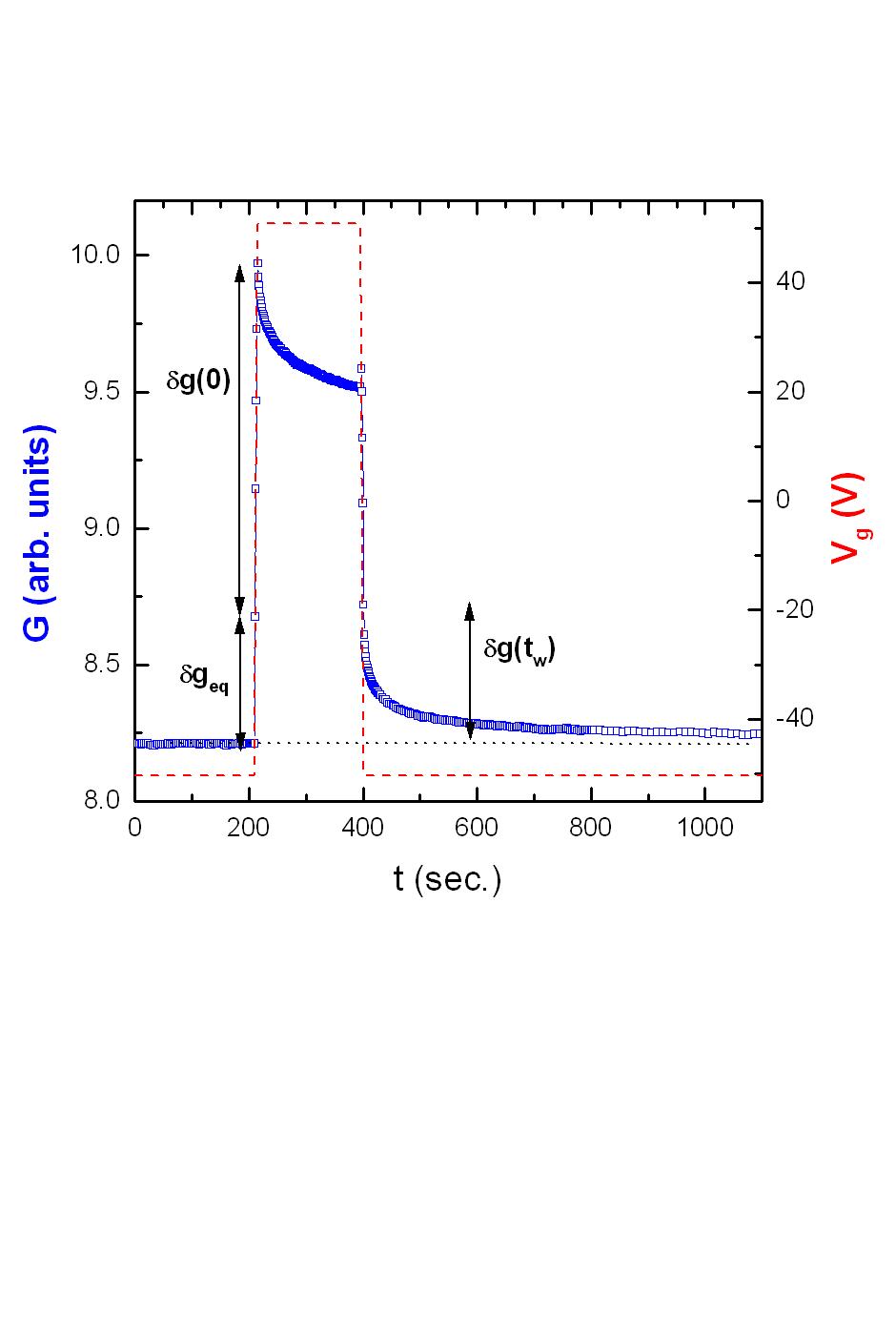}%
\caption{Typical run of the double-conductance-excitation protocol. The
dependence of the conductance $G$ on time is plotted as squares and the gate
voltage as dashed line. The sample has $R_{\square}=22.5M\Omega$ is
$30\mathring{A}$ thick and 1x1mm lateral dimensions, and measured under
$F=30V/cm.$ A constant waiting time $t_{w}=3\min$ has been used for all
experiments in this work. The bath temperature is $T=4.11K$.}%
\end{center}
\end{figure}
Starting with the sample with a voltage $V_{g}^{o}$ held at the gate, and
equilibrated under the fixed external conditions (temperature $T$, or electric
field $F$), one monitors the conductance as function of time $G(t)$ to obtain
a baseline conductance. Then, the gate voltage is switched to $V_{g}^{n}$ and
is maintained there for a certain \textquotedblleft
waiting-time\textquotedblright\ $t_{w}$. Finally, the gate voltage is switched
back to $V_{g}^{o},$ and $G(t)$ is measured for an additional period of time.
Here we are interested in the ratio $\delta g(0)/\delta g(t_{w})$ which is a
measure of relaxation time defined for \textit{a fixed} $t_{w}$. Actually,
this procedure is just the gate-protocol for aging described elsewhere
\cite{4}. The difference is that in the aging protocol one varies $t_{w}$
under fixed external conditions (i.e., $T$ or $F$) whereas here $t_{w}$ is
fixed and we vary $T$ (or $F$). A constant value $t_{w}=180\sec$ was used for
all the experiments reported here.

To see why $\delta g(0)/\delta g(t_{w})$ is related to dynamics note that
$\delta g(t_{w})$ is a measure of how far the sample $G$ has drifted in
phase-space during $t_{w}~$towards its equilibrium state (set by $V_{g}^{n}$
and the external conditions). If, for example, a full equilibrium is reached
during $t_{w}$, $\delta g(t_{w})$ will obviously equal $\delta g(0).$ If, on
the other hand, relaxation is infinitely slow, $\delta g(t_{w})$ will be zero.
The initial conductance jump $\delta g(0)$ is a proper normalization; when the
external conditions are changed, the degree of sample excitation (when
$V_{g}^{n}$ is switched to $V_{g}^{o}$ or \textit{vice versa}) will in general
change too, and it will be reflected in $\delta g(0).$ Likewise, the value of
$\delta g_{eq}$ (c.f., figure 1) should be measured for each run because its
value usually depends on the external conditions. The origin of $\delta
g_{eq}$ is the equilibrium field effect, namely, the change of the equilibrium
conductance when $V_{g}^{n}$ is switched to $V_{g}^{o}~$(which in turn is due
to the energy dependence of the \textit{thermodynamic} density of states).
This physical quantity is reflected in an anti-symmetric contribution to the
field effect as illustrated in figure 2 for samples with different degrees of
disorder. The values of $\delta g(0)$ and $\delta g(t_{w})$ are extracted from
the $G(t)$ data as in figure 1 as follows. First, the times $t_{1}$ and
$t_{2}$ where $V_{g}^{n}$ reached its final value, and $V_{g}^{o}$ is
reinstated respectively are noted. These are used as the origin ($t=0)$ for
the two relaxations of $G(t);$ the first after the $V_{g}^{o}$ to $V_{g}^{n}$
switch, the second after the switch back. Each such $G(t)$ start out as
$G(t_{0})-a_{1,2}\log(t/t_{0})$ (the first, being history-free, persists as a
$\log$ throughout $t_{w},$ the second is logarithmic only for $t\ll t_{w}$).
The respective values of $G(t_{0})$ are found by extrapolation ($t_{0}$ is the
resolution time of the measurement, typically $t_{0}=1\sec$.), and are used to
calculate $\delta g(0)$ and $\delta g(t_{w})~$by subtracting the appropriate
baseline. This method should be easier to implement in systems such as
granular Al where the anti-symmetric part of $G(V_{g})$ is negligible relative
to the anomalous cusp in $G(V_{g})$ \cite{8} and thus there is no need for the
additional measurement of $\delta g_{eq}$.%
\begin{figure}
[ptb]
\begin{center}
\includegraphics[
trim=0.000000in 2.966654in 0.000000in 1.029970in,
natheight=7.083700in,
natwidth=5.000400in,
height=2.0747in,
width=3.3425in
]%
{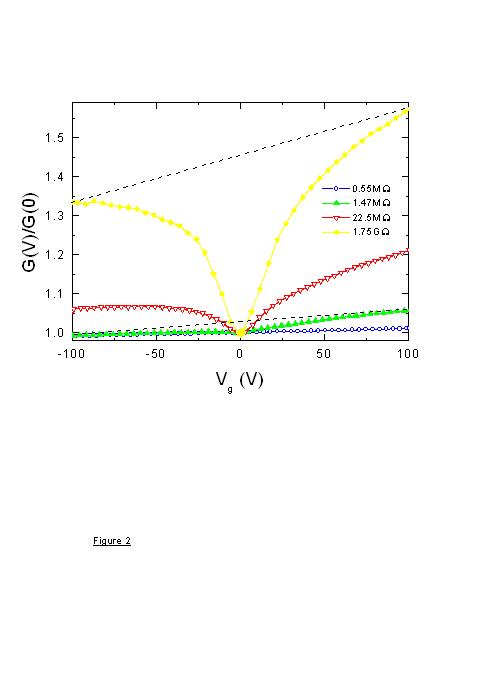}%
\caption{The normalized field effect at $T=4.11K$ for some of the samples used
in this work. Note that the relative amplitude of the anomalous cusp increases
with disorder, and is essentially zero for the sample with $R_{\square
}=0.55M\Omega.$ The dashed lines reflect the slopes of the equilibrium field
effect. These slopes obviously depend on disorder, as well as on temperature
and electric field (c.f., figures 4 and 5 respectively).}%
\end{center}
\end{figure}

\subsubsection{A single-conductance-excitation}

A quick way to assess dynamics is based on monitoring the relaxation of the
excess conductance created, say, by a $V_{g}$ switch. Then the resulting data
$G(t_{0})-a\log(t/t_{0})$ is normalized by $\delta G(t_{0})$ and the prefactor
$a/\delta G(t_{0})$ is used as a measure of the dynamics. This normalization
is not as reliable as the one used in the previous method because it does not
preclude the possibility that the excitation itself may depend on the external
conditions (namely, non-linear effects). Note also that unless $\delta g_{eq}$
is negligible it has to be measured separately and be used in the analysis.
This method will be used here only as a further illustration of the central
finding of our study.

\section{Results and discussion}

\subsubsection{Temperature dependence of the dynamics}

Measurements of glassy features as function of temperature are more complex
and demanding than it may be at first realized. This is mainly due to the
exponential temperature dependence of the conductance of electron glasses.
Note that a pre-requisite for observing non-ergodic effects in these systems
is that the sample is in the strongly localized regime. In two dimensions,
that means that $R_{\square}$ is larger than the quantum resistance
$h/e^{2}\approx25k\Omega.$ In fact, we observe glassy behavior only in samples
with $R_{\square}\geq200k\Omega,$ (see, for example, figure 2 and figure 4).
Such samples naturally exhibit exponential $G$ versus $T$. This exponential
sensitivity of $G(T)$ mandates a tight control over the bath temperature since
the glassy part of $G$ is of the order of only few percents. For the sample
used here it was necessary to establish a $\pm1mK$ temperature stabilization
at each instance of the studied range.

Another problem that makes these measurements a time consuming endeavour is an
inherent memory effect; Changing $T$ leaves the system with memory of the
previous conditions, and a long equilibration period is required before a
measurement at the new $T$ may commence \cite{9}. This effect will be
discussed and illustrated later (figure 11).

Once the bath temperature has stabilized on the target $T,$ a record of $G(t)$
was initiated to detect any drift that still exists due to the glassy
relaxation. The latter could be distinguished from the drift/fluctuation of
the bath temperature by the reading of a calibrated Ge-thermometer attached to
the sample holder. The measurements that are the basis for the results in
figure 3 included, at each temperature, a double-conductance-excitation
procedure as well as a $G(V_{g})$ scan such as that shown in figure 4 from
which the respective value of $\delta g_{eq}$ was obtained. The $G(t)$ data of
the conductance-excitation protocol was corrected for the spurious temperature
fluctuations by the respective trace of the Ge-thermometer. Finally, the
values of $\delta g(0)$ and $\delta g(t_{w})$ were calculated as described
above using the temperature-corrected$~G(t)$ in conjunction with $\delta
g_{eq}$ from the data in figure 4.%
\begin{figure}
[ptb]
\begin{center}
\includegraphics[
trim=0.000000in 6.076861in 0.000000in 2.118953in,
natheight=14.583300in,
natwidth=9.791400in,
height=2.0851in,
width=3.1808in
]%
{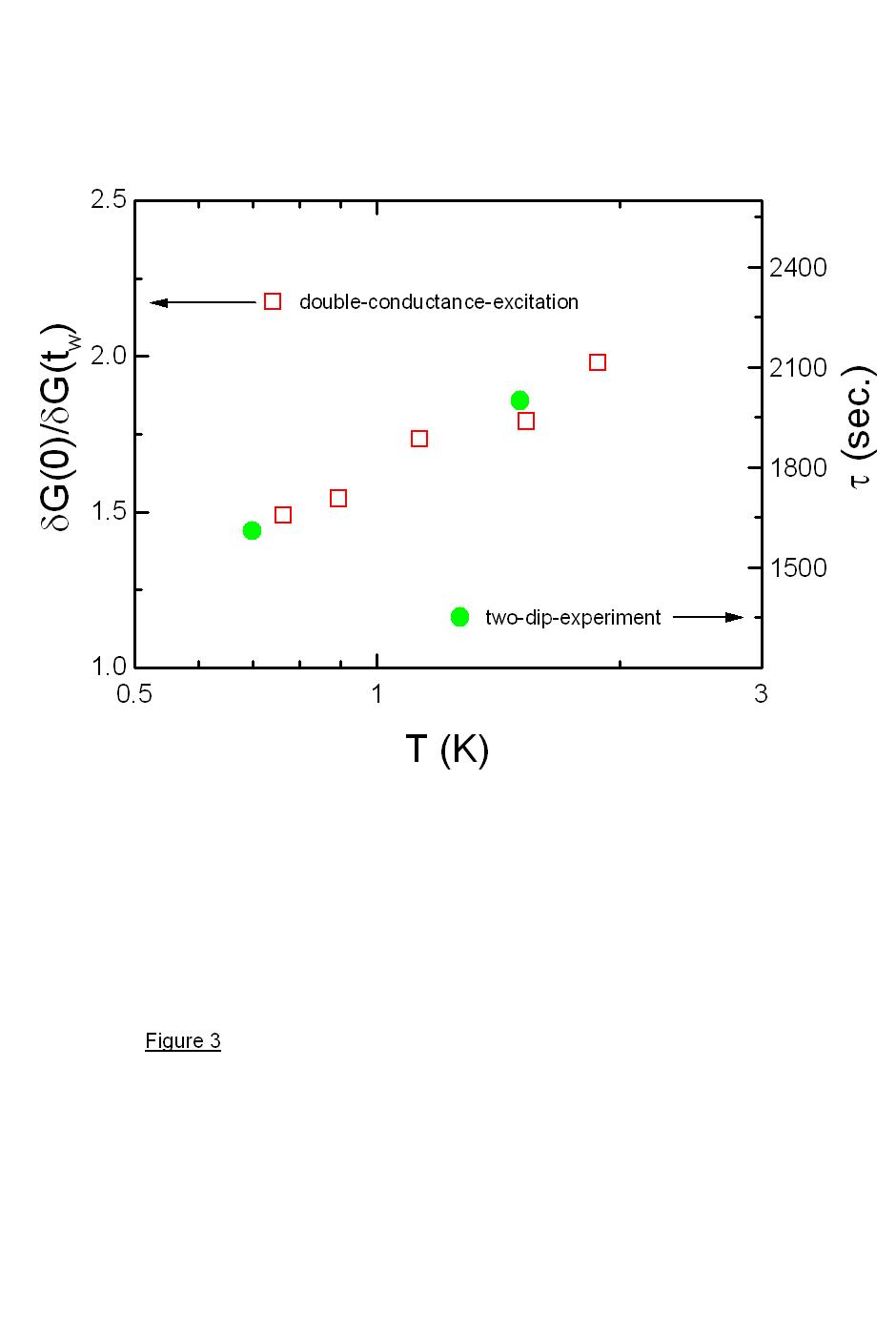}%
\caption{The dependence of the dynamics parameter $\delta g(0)/\delta
g(t_{w})$ on temperature. This is based on the double-conductance-excitation
using a constant $t_{w}=3\min$ (squares) and $\tau$ based on the
two-dip-experiment (circles). Data are for a $50\mathring{A}$ thick and 1x1mm
lateral dimensions sample with $R_{\square}=212k\Omega$ at $T=4.11K$.}%
\end{center}
\end{figure}

The results of the more elaborate two-dip-experiment performed at two
temperatures in the studied range are also shown in figure 3 for comparison.
Both methods yield the same result; the dynamics in this particular range of
disorder and temperature becomes somewhat \textit{slower} with $T$. The
dependence on temperature is rather weak; for a factor of 3 in the range of
$T,$ the average conductance changes by a factor of 25 while the
characteristic relaxation time $\tau$ changes by only 30\%. This slowing down
of the dynamics with temperature is consistent with a tunneling in a
dissipative environment scenario \cite{3}. We were not yet able to extend
these measurements to higher temperatures. In addition to the technical
difficulties noted above, the glassy effects on which the methods is based die
out exponentially with temperature. This can be clearly seen in the way the
anomalous cusp is diminished with $T$ (figure 4).%
\begin{figure}
[ptb]
\begin{center}
\includegraphics[
trim=0.000000in 4.580615in 0.000000in 2.123329in,
natheight=14.583300in,
natwidth=9.791400in,
height=2.5659in,
width=3.1808in
]%
{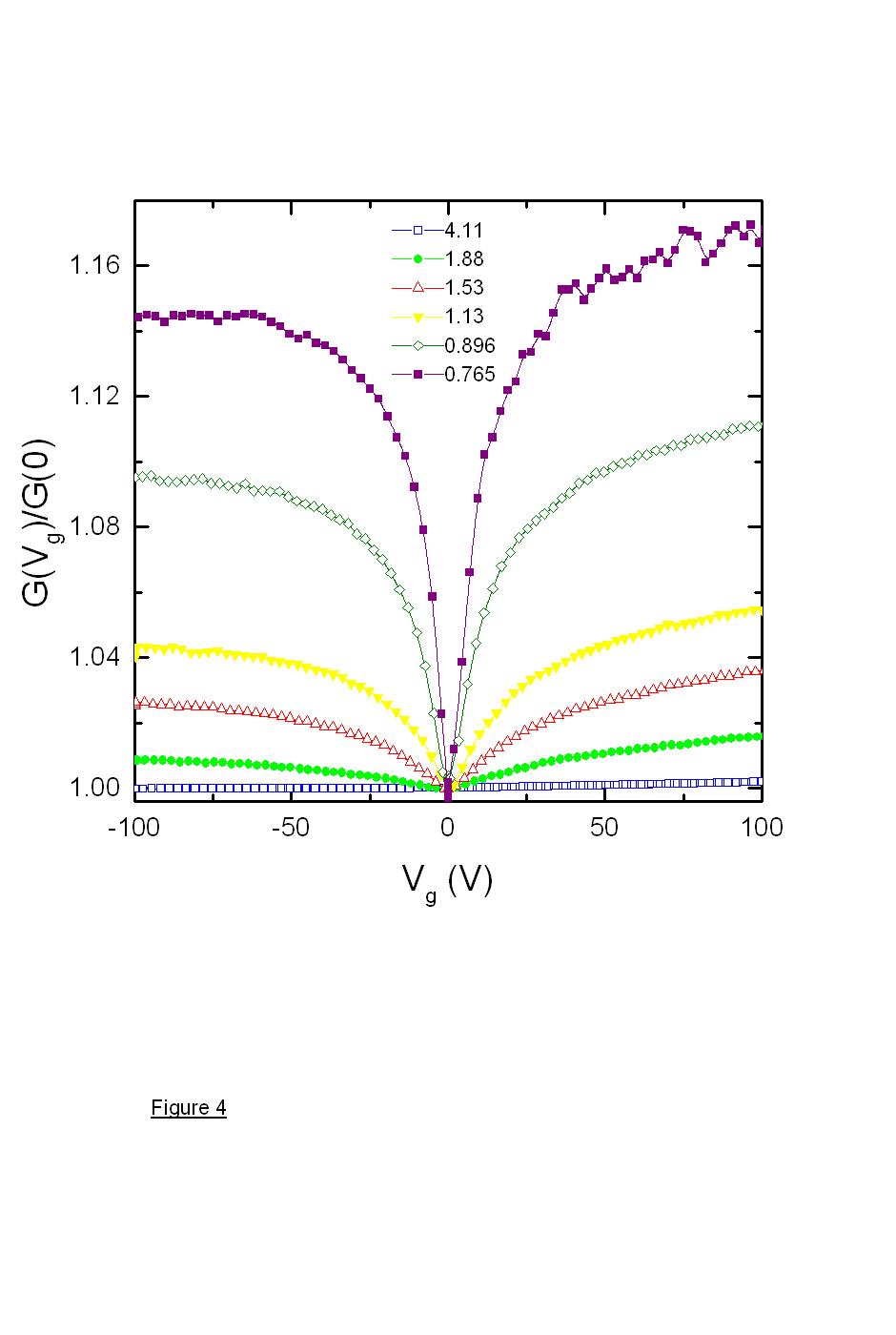}%
\caption{The normalized field effect at different temperatures for the same
sample as in figure 3. Note the dependence of the asymmetric component of the
field-effect on temperature as well as the fast diminishment of the anomalous
cusp amplitude with $T$.}%
\end{center}
\end{figure}
It should be noted that there is no a-priori reason that the observed trend in
$\tau(T)$ will change sign before the glassy effects become immeasurably
small. The transition to a glassy phase in the electron glass is a quantum
phase transition \cite{10}, and for a two-dimensional system there may be no
`glass temperature' except at $T=0$. Observation of glassy effects at a finite
$T$ would still be possible as long as $k_{B}T\eqslantless E_{C}$ (where
$E_{C}$ is the relevant interaction energy) in the same vein that insulating
features are observable in Anderson insulators at finite $T$ while, strictly
speaking, the metal-insulator transition occurs at $T=0$.

\subsubsection{Dependence on non-ohmic field}

Studying the dynamics as function of $F$ entails several advantages that make
the process easier than the respective study as function of $T$: Switching
from one value of $F$ to another is fast and accurate, the signal/noise
naturally improves with $F$ (as long as $F$ is much smaller than a critical
value see: \cite{11}), and working with a massive $^{4}$He Dewar as the base
temperature makes it feasible to perform long runs without thermally-cycling a
given sample. In the experiments detailed below, the equilibration times used
prior to running a double-conductance-excitation protocol was typically
several hours. It is important to realize though that even such an extended
period of equilibration does not guarantee that the measurement is performed
at equilibrium. In fact, relaxation from an excited state in these systems
were observed to persist for at least several days \cite{6}. As it is
impractical to wait that long at each field (or temperature), one has to
acknowledge the possibility that the results may deviate from their true
value. We did however test the sensitivity of our procedure to variation in
the equilibration time and found that, within the experimental error, waiting
more than 4 hours had a negligible effect on the outcome.

The same protocol was used here as in that leading to figure 3 above. For each
sample, and at each value of $F,$ a $G(V_{g})$ scan was taken to get the value
of $\delta g_{eq}.$ An example is shown in figure 5 for one of the samples.%
\begin{figure}
[ptb]
\begin{center}
\includegraphics[
trim=0.000000in 6.076861in 0.000000in 2.121870in,
natheight=14.583300in,
natwidth=9.791400in,
height=2.0833in,
width=3.1808in
]%
{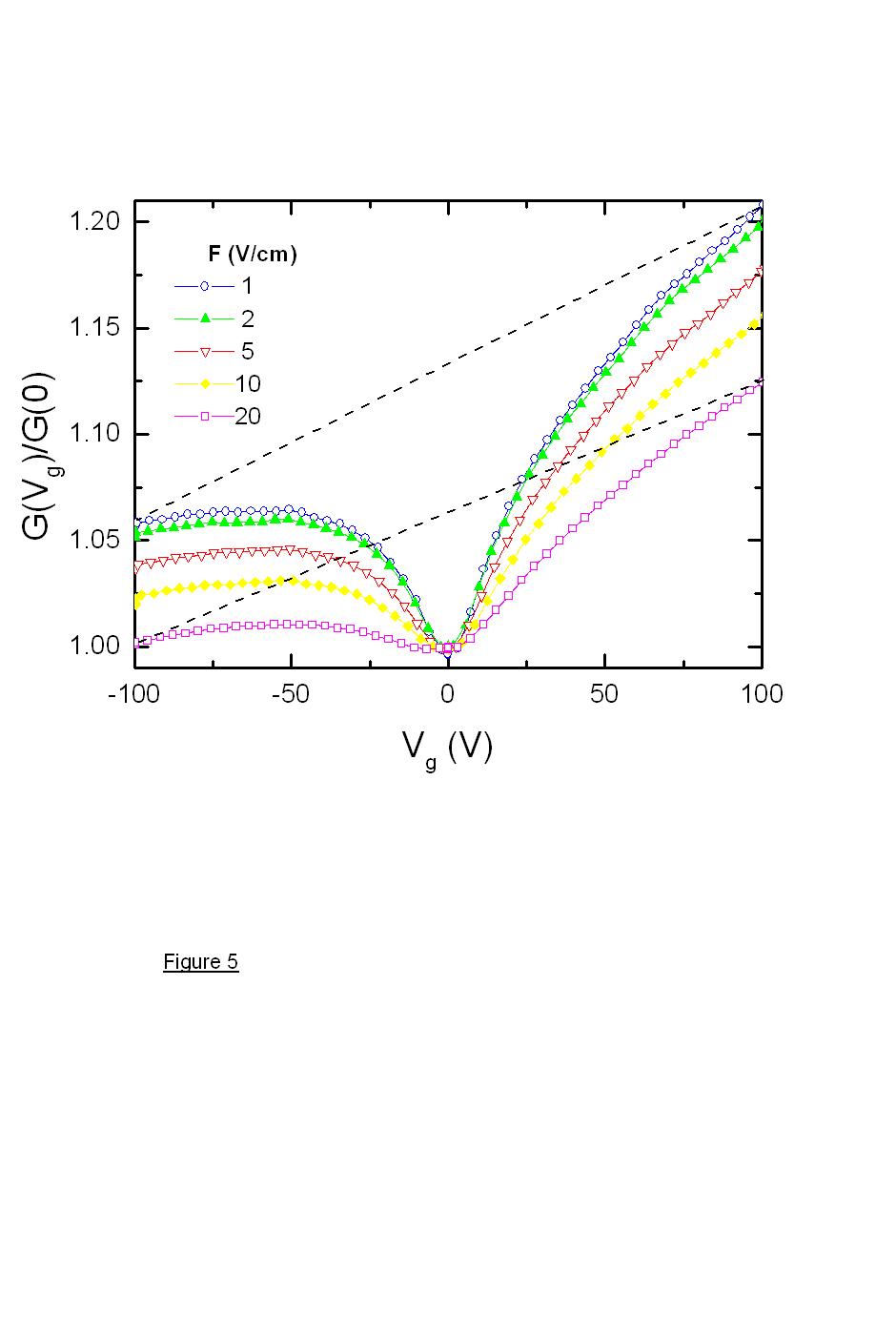}%
\caption{The normalized field effect measured at different values of the
electric field for the sample in figure 1. The dashed lines represent the
slopes of the equilibrium field effect. These are used to estimate the
relevant $\delta g_{eq}$ (see text).}%
\end{center}
\end{figure}

Figure 6 shows the dependence of the dynamics on $F$ for a $In_{2}O_{3-x}$
sample at three different states of disorder characterized by the sheet
resistance at $T=4.11K.$ These were produced from a single batch of
indium-oxide by UV-treatment as described elsewhere \cite{12}. Also shown
(lower graph) is the dependence of the conductance of these samples on the
field. Two observations can be made from the figure; first, for a given field,
$\delta g(0)/\delta g(t_{w})$ increases with disorder, which is consistent
with a previous study based on the two-dip-experiment \cite{7} (note that a
larger $\delta g(0)/\delta g(t_{w})$ means slower relaxation). Secondly,
$\delta g(0)/\delta g(t_{w})$ increases with $F,$ and there is an obvious
correlation between this dependence and the deviation from ohm's law of the
respective sample (lower graph). In particular, in the small field limit the
results for both $G(F)$ and $\delta g(0)/\delta g(t_{w})$ tend to become
independent of $F$ as one expects for the linear response regime.%
\begin{figure}
[ptb]
\begin{center}
\includegraphics[
trim=0.000000in 2.025620in 0.000000in 1.061664in,
natheight=14.583300in,
natwidth=9.791400in,
height=3.7282in,
width=3.1808in
]%
{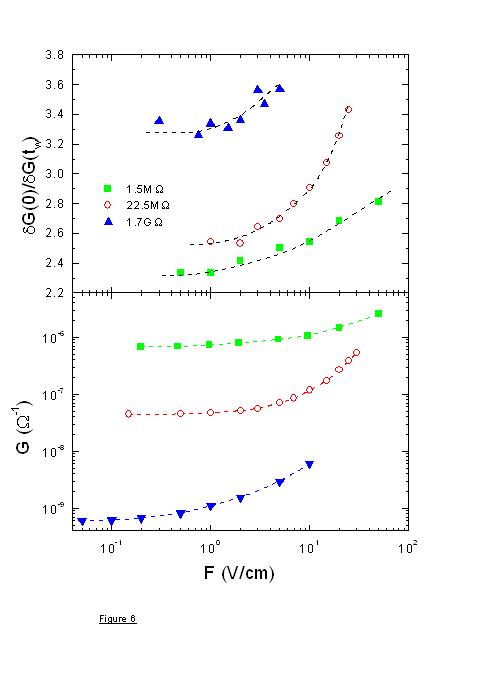}%
\caption{The dependence of the dynamics parameter $\delta g(0)/\delta
g(t_{w})$ on electric field for three samples with different disorder (upper
graph). Lower graph shows the dependence of the conductance of these samples
on field. The bath temperature is $T=4.11K$. Dashed lines are guide for the
eye.}%
\end{center}
\end{figure}

The increase of $\delta g(0)/\delta g(t_{w})$ with $F$ seems to mimic the
slowing-down with $T$ trend noted in figure 3 above. It is therefore tempting
to cast the results of the upper graph in figure 6 as function of some
effective temperature $T^{\ast}$. One way to do that is by combining the
$G(F)$ data with the conductance versus temperature $G(T)$ of these samples
exhibited in figure 7.%
\begin{figure}
[ptb]
\begin{center}
\includegraphics[
trim=0.000000in 3.428534in 0.000000in 2.028537in,
natheight=14.583300in,
natwidth=9.791400in,
height=2.9663in,
width=3.1808in
]%
{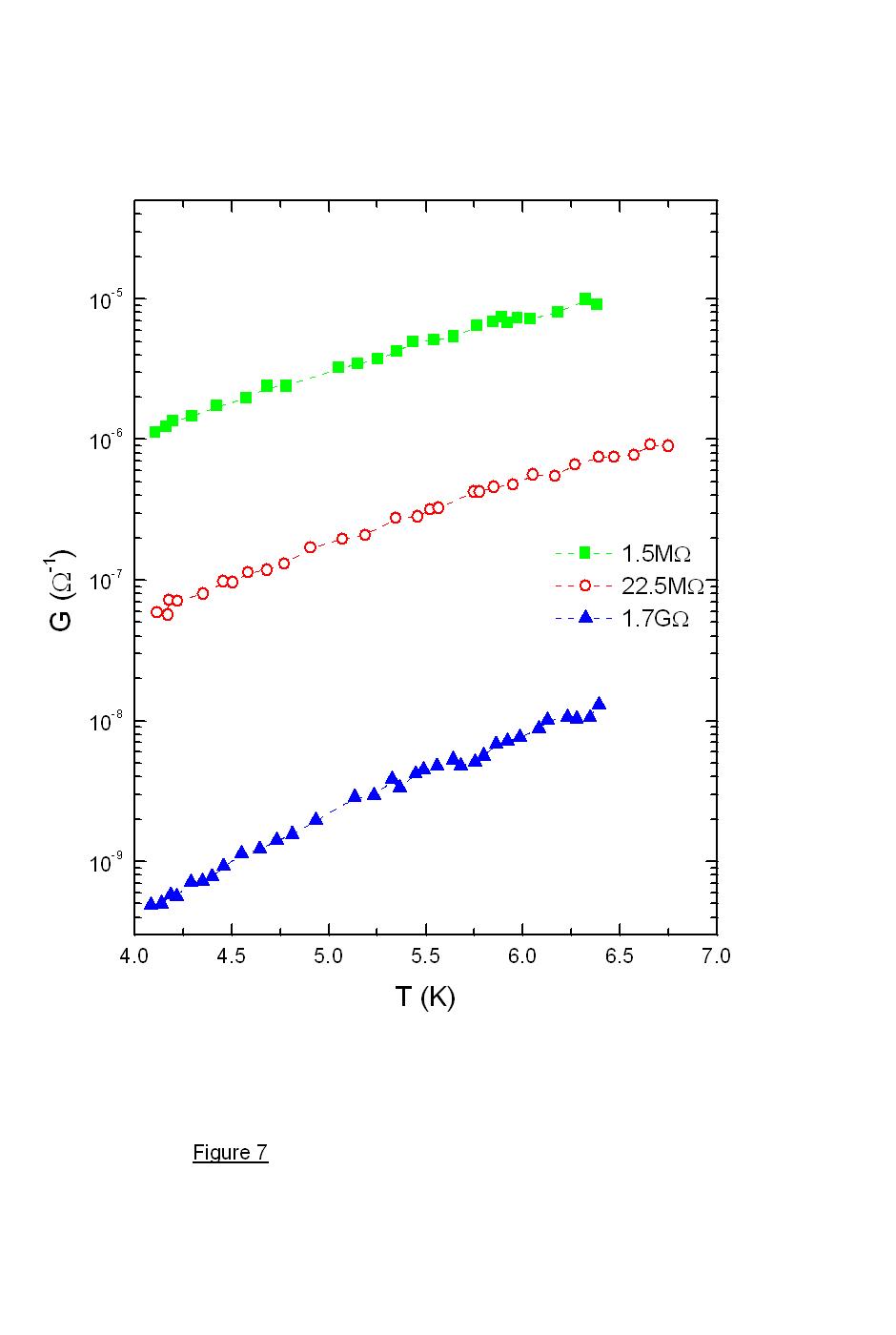}%
\caption{The dependence of the (ohmic) conductance $G$ on temperatures for the
samples in figure 6. These data were taken by raising the probe above the
liquid helium bath, then slowly lowering it back while monitoring $G$ and the
temperature by the reading of the Ge thermometer attached to the sample
stage.}%
\end{center}
\end{figure}
In essence, this procedure is tantamount to using the average conductance as a
thermometer. Reservations with this `effective temperature' approach will be
mentioned later. The results of this procedure are given in figure 8, which
for all three samples show a similar behavior of the dynamics as that
exhibited in figure 3 as function of temperature.%
\begin{figure}
[ptbptb]
\begin{center}
\includegraphics[
trim=0.000000in 6.078319in 0.000000in 2.120412in,
natheight=14.583300in,
natwidth=9.791400in,
height=2.0833in,
width=3.1808in
]%
{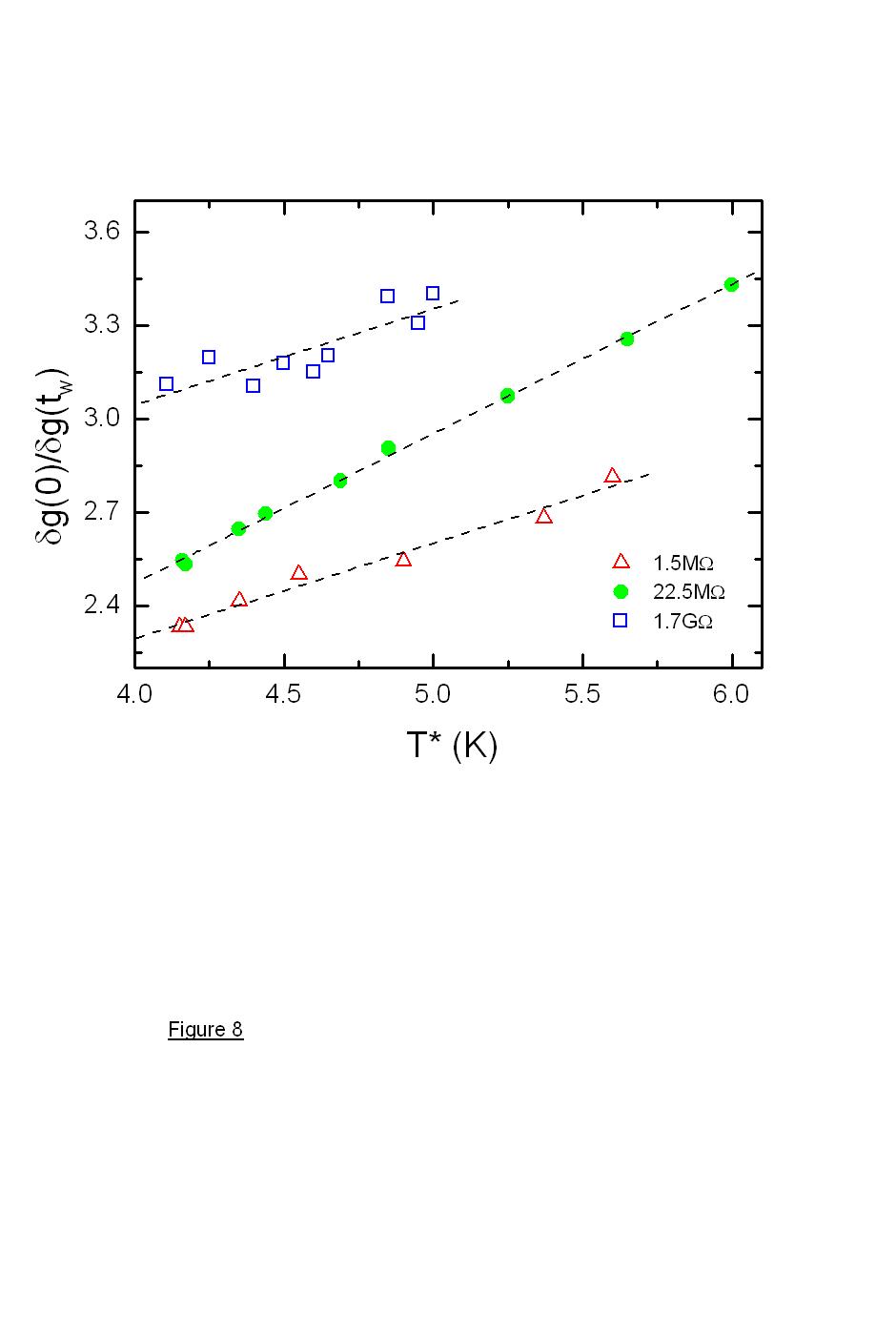}%
\caption{The dependence of the dynamics parameter $\delta g(0)/\delta
g(t_{w})$ on effective temperature $T^{\ast}$ using the data of figure 6 and
figure 7 (see text for details).}%
\end{center}
\end{figure}

The observation of a slower dynamics at higher temperatures is not a universal
feature of the electron glass. This dependence has been seen so far only in
crystalline indium-oxide films and over a limited range of temperatures. In a
preliminary study \cite{13} employing $50\mathring{A}$ thick $In_{2}O_{3-x}$
films doped with 3.4\% Au and compared their dynamics versus $F$ with that of
undoped $In_{2}O_{3-x}$ that otherwise had similar characteristics. Two pairs
of samples (Au doped and undoped) were studied so far with $R_{\square}$
$\approx1.5M\Omega$ and $\approx30M\Omega$ and the following results were
obtained: The undoped samples consistently showed virtually the same $F$
dependence as in figure 6. By contrast, over the same range of fields
($10^{-1}-10^{2}V/cm$), the two Au-doped samples exhibited much weaker $F$ -
dependent dynamics, in fact, within the experimental error, the dynamics was
independent of $F$. A possibly relevant difference between the two types of
materials appears to be a wider anomalous cusp which suggests a higher density
of carriers in the Au-doped samples \cite{14}. It would be interesting to see
if using a material with still higher carrier concentration will result in the
opposite trend of dynamics versus $T$. This may be attained for example with
amorphous indium-oxide films that can be prepared with a much wider range of
carrier concentration than can be achieved by doping $In_{2}O_{3-x}$ \cite{14}.

For the time being we shall focus on the dynamics of the crystalline samples
that exhibit dynamics slowdown with either $T$ or $F$. This unusual dependence
is interesting enough to justify further efforts to establish its existence
even for just its uniqueness; we are not aware of any other glass that
exhibits this kind of behavior. Moreover, it would seem that such a dependence
might be reconciled with quantum relaxation making the electron glass a prime
candidate for such studies.

In the following we describe two experiments demonstrating qualitatively that
the dynamics of a crystalline $In_{2}O_{3-x}$ slows down upon an increase in
the non-ohmic $F.$ In the first experiment, we used the
single-conductance-excitation method, and generated normalized relaxation
curves measured under different non-ohmic fields. These are plotted in figure
9 along with the resistance versus field characteristics of this sample. The
slowing down of the dynamics is clearly reflected in the logarithmic slopes of
the relaxation curves becoming smaller as $F$ increases. (Note however that
the values of $\delta g_{eq}$ were not excluded from the data in this case,
and therefore the effect is somewhat smaller than that reflected in the
figure).
\begin{figure}
[ptb]
\begin{center}
\includegraphics[
trim=0.000000in 0.756539in 0.000000in 0.454773in,
natheight=7.083700in,
natwidth=5.000400in,
height=3.9211in,
width=3.3425in
]%
{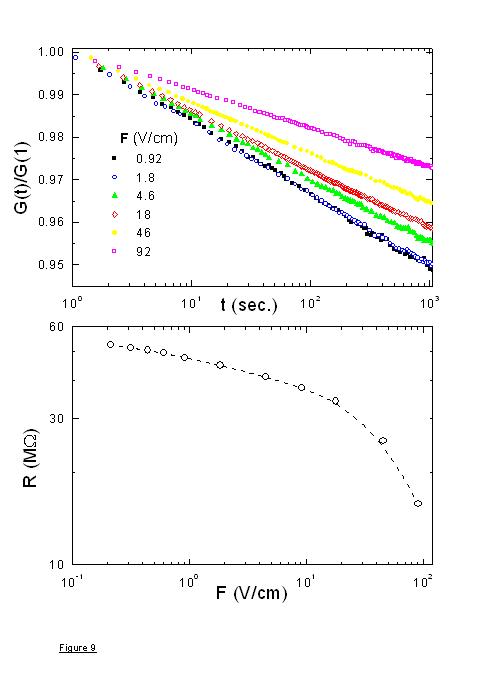}%
\caption{Results of a single-conductance-excitation protocol at various
electric fields. Sample is $52\mathring{A}$ thick, with 1.1x1mm lateral
dimensions and has $R_{\square}=52.5M\Omega$ at $T=4.11K$ (in the limit of
small field). Note that the logarithmic slopes get progressively smaller when
$F$ gets further into the non-ohmic regime (This can be seen in the lower
graph that depicts the resistance versus $F$ of this sample).}%
\end{center}
\end{figure}

The second experiment, utilizing the same sample, is the analog of the
temperature excursion protocol (referred to sometimes as `rejuvenation' in the
dielectric-glass and spin-glass communities \cite{15}). In the experiment,
shown in figure 10, the sample is initially relaxing from an excited state
(produced by a quench-cool or gate switch), and its conductance versus time is
recorded for $t_{1}\approx100\sec$ while under a relatively high $F_{1}$.
Then, $F_{1}$ is suddenly switched to a near-ohmic field $F_{2}$, and $G(t)$
(naturally now lower than when under $F_{1}$) is measured for $t_{2}%
\approx1300\sec$ before $F_{1}$ is re-instated, and the relaxation of $G$ is
extended for $t_{3}\approx$ $27,000\sec$. Note that the $G(t)$ during $t_{3}$
is a natural extension of the purely logarithmic relaxation of the $t_{1}$
period, which is the natural relaxation law of the electron glass \cite{5,6}.
However, to get the $G(t)$ in these two regions to match as well as in the
figure, the time axis of the $t_{3}$ region was \textit{right-shifted} by
$\approx400\sec.$This is precisely the opposite direction that is necessary to
match the respective relaxations in the `classical' glasses where the dynamics
slows down upon cooling \cite{15}. Note also that the logarithmic slope of the
relaxation during $t_{2}$ is consistent with a \textit{faster} dynamics at the
lower `effective $T$'.%
\begin{figure}
[ptb]
\begin{center}
\includegraphics[
trim=0.000000in 5.983528in 0.000000in 2.027079in,
natheight=14.583300in,
natwidth=9.791400in,
height=2.1447in,
width=3.1808in
]%
{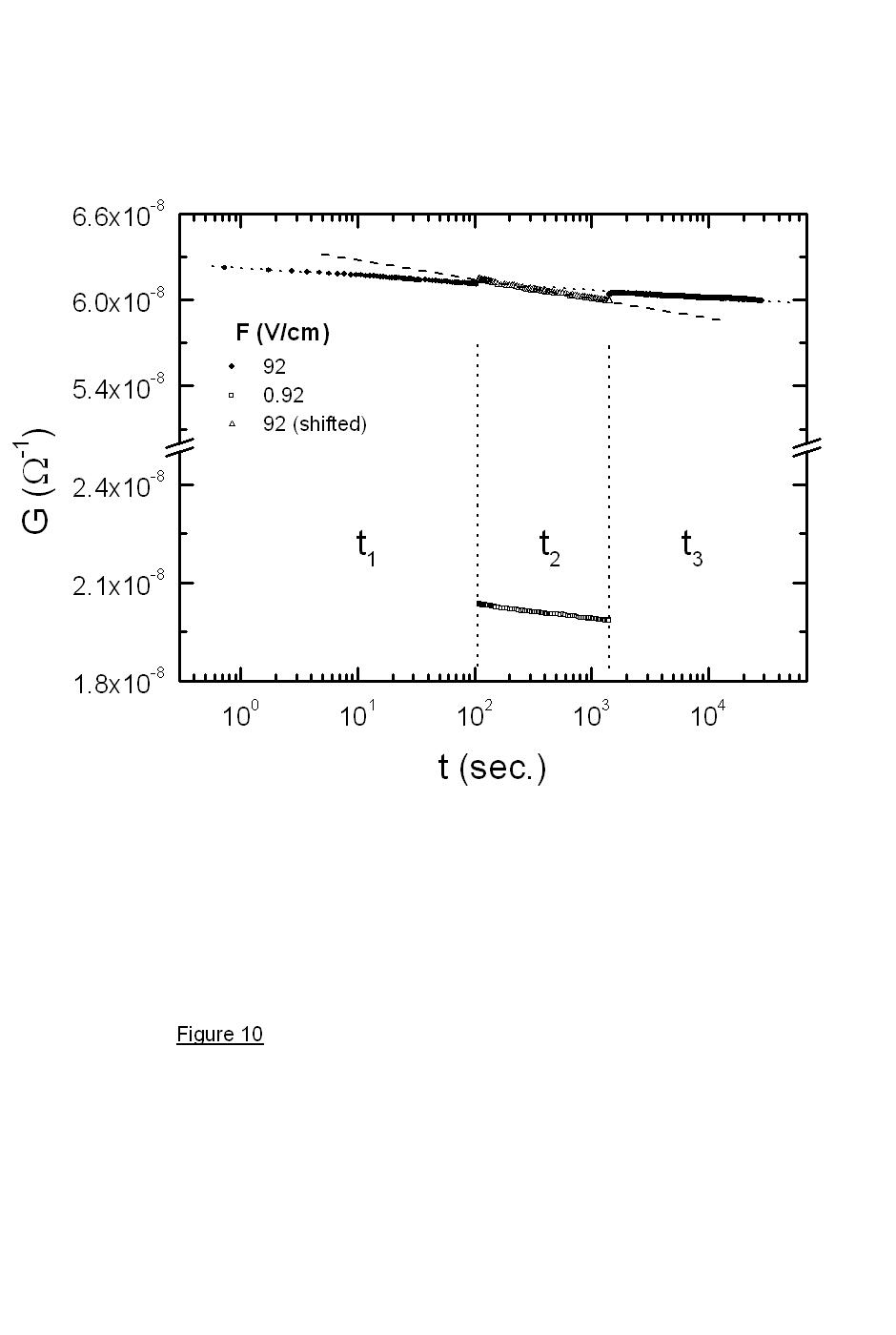}%
\caption{The conductance versus time during a field-cycling protocol as an
analog of a `rejuvenation' experiment (see text). The same sample as in figure
9.}%
\end{center}
\end{figure}

\subsubsection{Is a non-ohmic field equivalent to a higher `effective'
temperature?}

As explained above, using high fields has some technical advantages over
raising the temperature. The question however is to what degree the effects
produced by a large $F$ are the same as those produced by raising $T$. In
general, the state of the system under large \thinspace$F$ would be different
than that achieved by a higher $T.$ This is true in the diffusive regime
\cite{16}, in the hopping regime, and whether or not the system is glassy. One
may be assured that `heating' and `non-equilibrium' are indistinguishable only
if the relevant distribution functions assume their \textit{equilibrium} form
and with the \textit{same} temperature. In terms of measured quantities
however, the difference between heating and a high-field situation may be
small when the various scattering lengths (for momentum, and energy
relaxations) are much smaller than the sample dimensions. The samples used
here (typically, 1x1mm) are much larger than the largest scale in the problem
(which is the percolation radius that at the lowest $T$ used here is of order
$10^{3}$\AA , the hopping length is one order of magnitude smaller). It is
plausible that, under these conditions, measured quantities will be affected
in the same direction by increasing either $F$ or $T$, and this is the only
assumption made here. We do not expect that the mapping used in arriving at
$T^{\ast}$ in figure 8 should be identical to a measurement using an
equilibrium bath-temperature $T~$that equals $T^{\ast}$. In fact, as the next
experiment shows, warming up the sample to obtain the same $G$ as that
resulting by a non-ohmic $F$ may lead to a \textit{quantitatively} different effect.

Figure 11 shows the time dependence of the conductance of a sample under two
different external conditions; Starting from a near-equilibrium state at
$T=4.1K$ the sample was quickly warmed up to $T=6.2K,$ which caused its
(ohmic) conductance to increase by a factor of $\approx3.$ It was then kept at
this $T$ while its $G(t)$ was recorded for $\approx1000\sec.$ Next, the sample
was cooled back to $T=4.1K,$ allowed to equilibrate for 24 hours, and then it
was subjected to a field $F$ such that the conductance measured immediately
after the switch to the non-ohmic field has increased to the same value
($\pm2\%)$ as the respective $G$ immediately after the switch to $T=6.2K~$(see
figure). Then $G(t)$ was measured while holding this $F$ at a bath $T=4.1K.$
The ensuing $G(t)$ increases with time in both cases, however the $T=6.2K$
variant increases noticeably faster, and it may have a somewhat different
functional dependence than the non-ohmic $F$ time-trace.

The logarithmic increase of $G$ with time under the non-ohmic field has been
studied before as a manifestation of a time-dependent heat-capacity \cite{17}.
The difference between the effect of field versus that of temperature is not
surprising; The field coupling to the hopping system is both inhomogeneous and
anisotropic whereas temperature (via phonons) couples effectively to all
regions of the sample. Aiming to set $G$ (at a given time) at a predetermined
value by raising $T$ will generally entail quite different current-carrying
network than when the same goal is accomplished by $F$. This difference in
also relevant for the energy-balance that determines the steady-state
conductance (reached for $t\geq t_{ergodic}$), which therefore will in general
be different for the two routes.%
\begin{figure}
[ptb]
\begin{center}
\includegraphics[
trim=0.000000in 5.922278in 0.000000in 2.121870in,
natheight=14.583300in,
natwidth=9.791400in,
height=2.1335in,
width=3.1808in
]%
{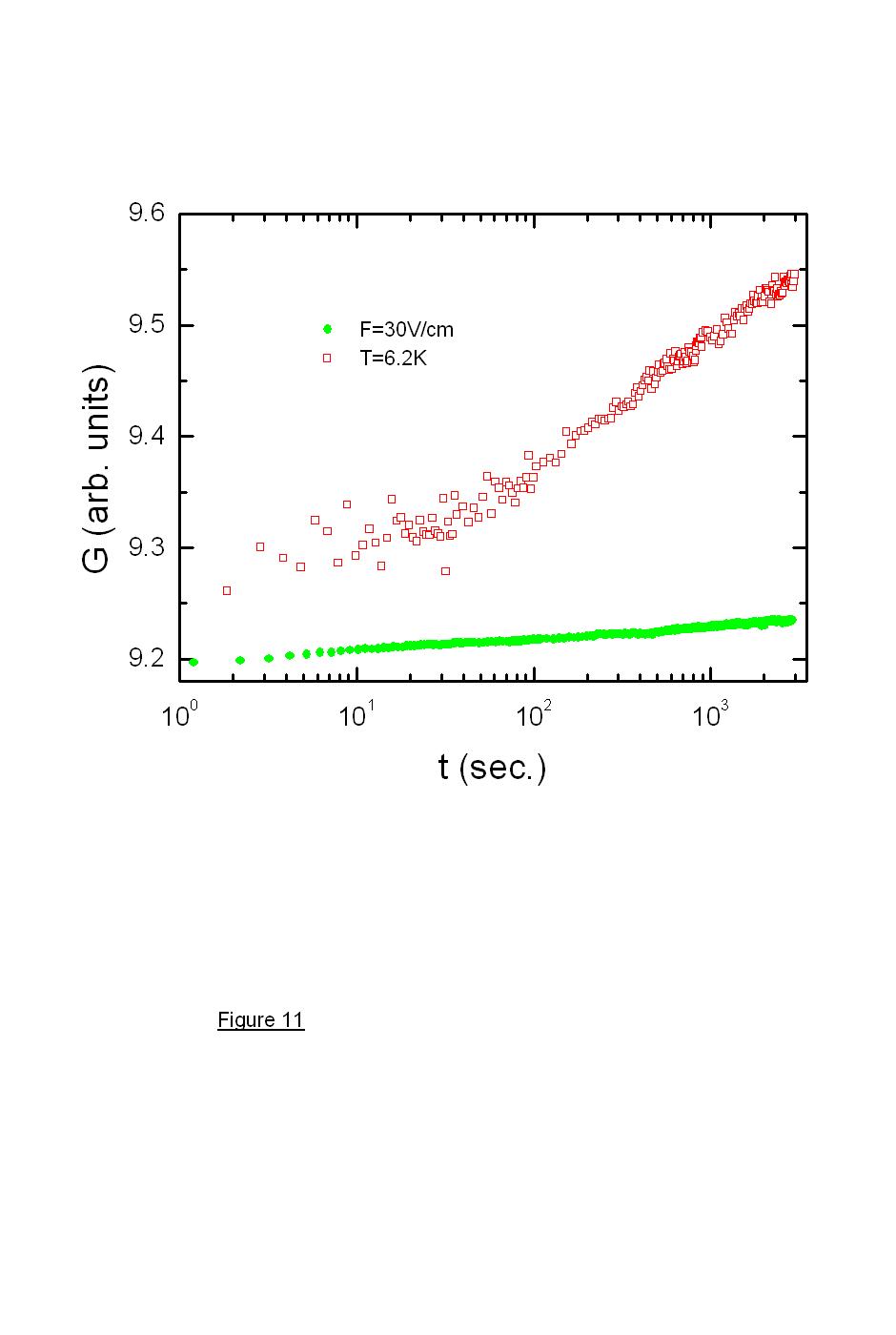}%
\caption{Conductance versus time after quickly raising the temperature of a
sample ($50\mathring{A}$ thick, $R_{\square}=1.42M\Omega$ at $T=4.11K$ where
it equilibrated for $t\eqslantgtr20$ hours) to $T=6.2K$ (squares), and after
incraesing the field from its ohmic value to $F=30V/cm$ which gave the almost
the same $G$ while keeping the sample at the liquid helium bath (circles).}%
\end{center}
\end{figure}

The next experiment carries a double message; It is yet another illustration
that $F$ and $T$ produce qualitatively similar effects, and it demonstrates
that the concept of `effective temperature' has at best a limited value for
the glass state. The first half of this experiment shown in the lower graph of
figure 12 is the analog of the experiment described before involving a
temperature quench (see figure 10 in reference 5). Starting with the sample
equilibrated for 23 hours under $V_{g}=50V$ and $F_{1}=92V/cm$ a field effect
scan $G(V_{g})$ is taken to show the shape of the near-equilibrium anomalous
cusp that characterizes these conditions. Then, $F_{1}$ is quickly reduced to
$F_{2}=0.45V/cm$ and a $G(V_{g})$ scan is taken consecutively at different
times after the switch to $F_{2}.$ Note that the sample conductance changes in
the $F_{2}\rightarrow F_{1}$ switch by $\approx320\%$ and this change occurs
\textit{immediately} after $F_{1}$ is set ($G$ continues to change
logarithmically after the "quench" by only $\approx2$ \% over the next 25
hours, c.f., figure 12). By contrast, the $G(V_{g})$ obtained a short time
after the switch to $F_{1}$ is essentially identical to the equilibrium result
under $F_{2}$! In other words, the system retains a memory of the value of the
field at which it was allowed to equilibrate. This observation, and the
eventual forming of the more prominent and sharper dip \cite{17} appear to be
a replica of the $T_{2}\rightarrow T_{1}$ experiment described elsewhere
\cite{5}. At the same time, note that, immediately after the $F_{2}\rightarrow
F_{1}$ switch one faces an ambiguous situation; Reading the value of $G$ one
concludes that the system is already under $F_{1}$. However, judging by the
shape of $G(V_{g})$ (which is indistinguishable from its steady-state shape
under $F_{2}$), one may just as well conclude that the system is still under
$F_{2}$. These experiments then demonstrate that two `thermometers' attached
to the same system (namely, the \textit{value} of $G(T^{\ast}),$ and the
\textit{shape} of the function $G(V_{g},T^{\ast}$) may give quite different
readings. This is one of the earmarks of a far-from-equilibrium situation
\cite{16}.%
\begin{figure}
[ptb]
\begin{center}
\includegraphics[
trim=0.000000in 1.994995in 0.000000in 1.090831in,
natheight=14.583300in,
natwidth=9.896100in,
height=3.7308in,
width=3.2136in
]%
{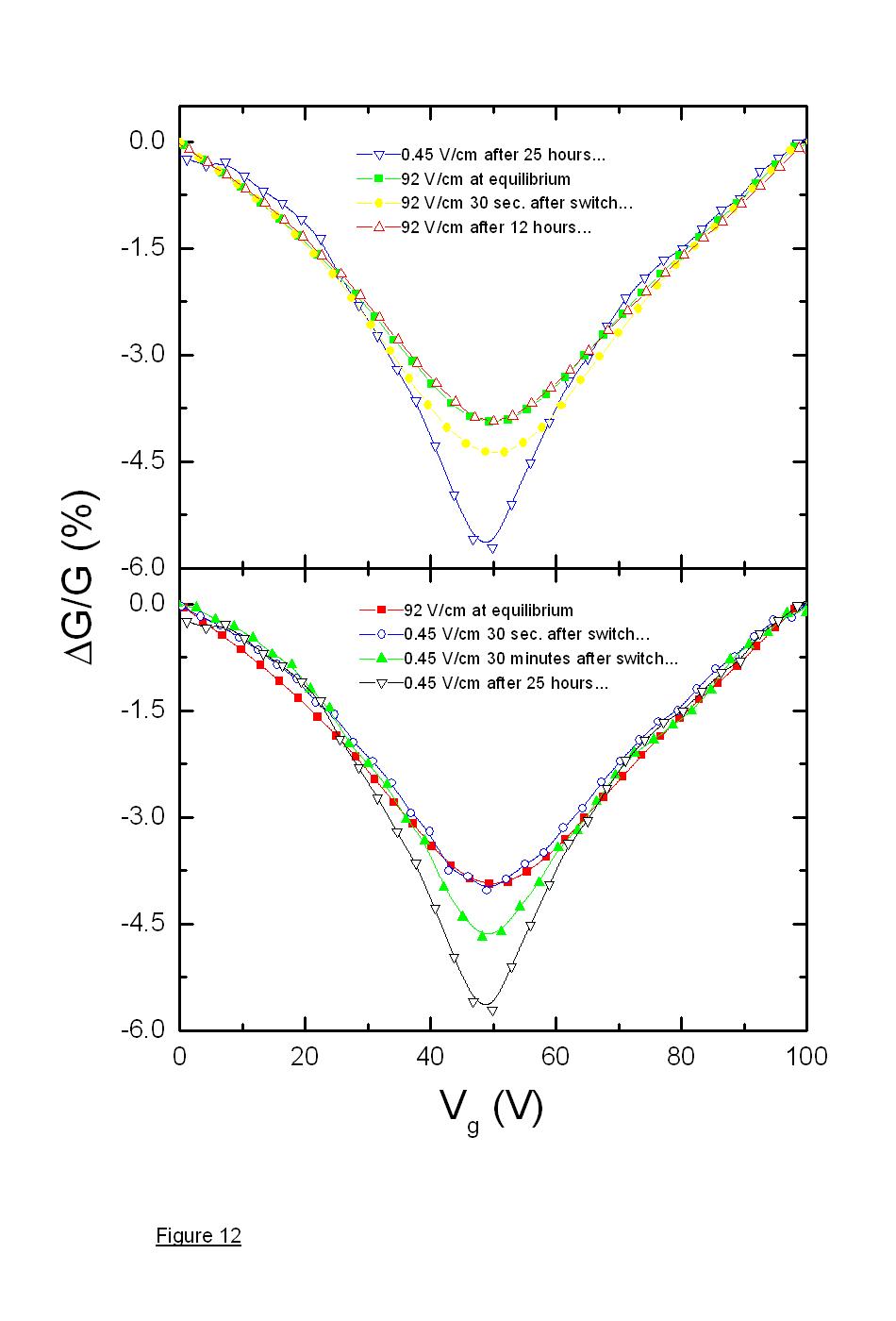}%
\caption{Memory of a previous state experiment switching between a relatively
high field ($92V/cm$) and an near ohmic field ($0.45V/cm$). The figure shows
the anomalous field effect (substracting the antisymmetric component from the
$G(V_{g})$ scan), under various conditions. Lower graph: Starting from
equilibrium under $92V/cm$ and switching to $0.45V/cm$. Upper graph: Starting
from immediately after switching back from $F=0.45V/cm$ to $92V/cm$. Same
sample as in figure 9.}%
\end{center}
\end{figure}

The current version of this `memory-of-a-previous-state' employing $F$ instead
of $T$ has the advantage of the speed and accuracy alluded to above. This also
makes it feasible to carry out the complementary procedure - go from the
near-equilibrium `cold' state to the `hot' state that, for technical reasons,
was not feasible with the $T_{2}\rightarrow T_{1}$ experiment \cite{18}. Note
(upper graph in figure 12) that following the $F_{1}\rightarrow F_{2}$ switch,
the sharper appearance of the dip that characterizes its shape at the low
field is essentially lost, and the shape of $G(V_{g})$ is much closer to the
near-equilibrium shape eventually attained than in the respective situation of
the complementary case. Note however that the conductance immediately after
the switch is still somewhat lower than its asymptotic value, which is
consistent with the time-evolution of the conductance in figure 11 (see also
reference \cite{19}). This illustrates again that changing $F$ (or $T$) in
either direction is accompanied by a two-stage conductance change; an initial
fast change $\delta G_{f}$ followed by a sluggish change $\delta Gs$ that has
the same sign as $\delta G_{f}$. Naturally, this effect should be taken into
account when analyzing experiments such as that in figure 10. The asymmetry in
the resulting effects of `cooling' versus `heating' presumably reflects the
difference between relaxation and excitation; relaxation is slow while
excitation is fast (see, for example figure 1).

In summary, we have described a number of experimental procedures designed to
probe the dynamics of electron glasses as function of temperature and fields.
These methods, applied to crystalline indium-oxide films consistently exposed
a non-trivial dynamics. Namely, the dynamics associated with relaxation
processes became more sluggish with temperature or when measured using
non-ohmic fields. It is tentatively conjectured that such a behavior reflects
the dynamics expected of a quantum glass where diffusion in phase space is
controlled by tunneling in the presence of dissipation \cite{3}. Experiments
on other type of materials are planned to further check on this conjecture. At
any rate, that dynamics which slows down with temperature is attainable at
some range of parameters must be viewed as a constraint on any model for the
electron glass.

This research was supported by a grant administered by the US Israel
Binational Science Foundation and by the Israeli Foundation for Sciences and Humanities.


\begin{thebibliography}{99}
\bibitem {9}In practice, one waits for the system to relax as long as $\Delta
G(t)$ during the planned experiement is smaller than the resolution. This
favors making small steps in $\Delta T$.
\bibitem {19}The experiment in reference 5 employed pumped helium cryostat.
Warming up from $T=1.6K$ back to $4.11K$ took much longer than typical
relaxation time of the electron glass while cooling down could be done fairly quickly.
\end{thebibliography}
\end{document}